\documentclass[12pt,journal,onecolumn]{IEEEtran}
\usepackage{amsmath,amssymb,amsfonts,dsfont,footnote}
\usepackage{pdfsync}
\usepackage{subfigure}
\usepackage{array}
\usepackage{graphicx}
\usepackage{mathrsfs}
\usepackage{cite}

\newtheorem{theorem}{Theorem}
\newtheorem{definition}{Definition}
\newtheorem{remark}{Remark}
\newtheorem{lemma}{Lemma}
\newtheorem{example}{Example}

\newcommand{\xH}{\mathbf{H}}
\newcommand{\xX}{\mathbf{X}}

\newcommand{\xZ}{\mathbf{Z}}

\newcommand{\sA}{\mathcal{A}}
\newcommand{\sB}{\mathcal{B}}
\newcommand{\sF}{\mathcal{F}}

\newcommand{\xxZ}{\mathds{Z}}
\newcommand{\xxK}{\mathcal{K}}
\newcommand{\xxM}{\mathcal{M}}
\newcommand{\xxN}{\mathcal{N}}

\newcommand{\xxL}{\mathcal{L}}
\newcommand{\SNR}{\mathrm{SNR}}
\newcommand{\dof}{\mathsf{DoF}}
\newcommand{\dofu}{\overline{\mathsf{DoF}}}
\newcommand{\dofl}{\underline{\mathsf{DoF}}}
\newcommand{\define}{\triangleq}

\long\def\symbolfootnote[#1]#2{\begingroup%
\def\thefootnote{\fnsymbol{footnote}}\footnote[#1]{#2}\endgroup}
\begin{document}
\title{Interference Alignment for the $K$-User MIMO Interference Channel}

\author{ Akbar Ghasemi,~Abolfazl~Seyed~Motahari,~and~Amir~Keyvan~Khandani
\\\small Coding \& Signal Transmission Laboratory (www.cst.uwaterloo.ca)
\\\small Department of Electrical and Computer Engineering,  University of Waterloo
\\\small Waterloo, ON, Canada N2L3G1
\\ \{aghasemi, abolfazl, khandani\}@cst.uwaterloo.ca
}

\maketitle

{\let\thefootnote\relax\footnotetext{Financial supports provided by Natural Sciences and Engineering Research Council of Canada (NSERC) and Ontario Ministry of Research \& Innovation (ORF-RE) are gratefully acknowledged.}}

{\let\thefootnote\relax\footnotetext{The material in this paper was presented in part at
the IEEE International Symposium on Information Theory (ISIT), Austin, Texas,
June 2010.}}

\begin{abstract}
We consider the $K$-user Multiple Input Multiple Output (MIMO)
Gaussian interference channel with $M$ antennas at each transmitter
and $N$ antennas at each receiver. It is assumed that channel
coefficients are constant and are available at all transmitters and at
all receivers. The main objective of this paper is to characterize
the Degrees of Freedom (DoF) for this channel. Using a new
interference alignment technique which has been recently introduced
in \cite{abolfazl-final}, we show that $\frac{MN}{M+N} K$ degrees of
freedom can be achieved for almost all channel realizations. Also, a
new upper-bound on the DoF of this channel is provided. This
upper-bound coincides with our achievable DoF for $K\geq
K_u\define\frac{M+N}{\gcd(M,N)}$, where
$\gcd(M,N)$ denotes the greatest common divisor of $M$ and $N$. This
gives an exact characterization of DoF for $M\times N$ MIMO Gaussian
interference channel in the case of $K\geq K_u$.
\end{abstract}

\newpage

\section{Introduction}
\PARstart{I}{nterference management} is one of the main challenges
in wireless networks in which multiple transmissions occur
concurrently over a common medium. Interference is usually handled
in practice either by \emph{interference avoidance}, in which users
coordinate their transmissions by orthogonalizing their signals in
time or in frequency, or by \emph{treating-interference-as-noise},
in which users adjust their transmission power and treat each
other's interference as noise. \emph{Interference decoding},
although more demanding, is another approach in which interference is decoded along with the desired signal.

During the past three decades, information theorists have made
extensive efforts to characterize the impact of the interference on
the capacity of wireless networks. For the two-user Gaussian
Interference Channel (IC), the capacity region has been
characterized for some ranges of channel coefficients
\cite{Carl,Han-Kob,Sason,SKC,abolfazl-inter,Annap}. For the general two-user case, a
characterization of the capacity region within one bit has been
presented in~\cite{ETW}.

By moving from the two-user case to \emph{more than two users}, the
capacity characterization becomes more challenging. To reduce the
severe effect of the interference for $K>2$ users, the use of a new
technique known as \emph{interference alignment} is essential.
Interference alignment, which was first introduced by Maddah-Ali
\emph{et al.} \cite{maddahali-isit,maddahali2008com} in the context
of MIMO $X$ channels, is an elegant technique that reduces the effect of the aggregated interference from several users to that of a single user
. This is accomplished by assigning a portion of the available time/frequency/space at each
receiver to the interference and enforcing all the interfering terms
to be received in that portion. There are two versions of
interference alignment in the literature: \emph{signal space alignment}
and \emph{signal scale alignment}. In signal space alignment, the transmit
signal of each user is a linear combination of some
vectors where data determines the coefficients of this linear combination.
In this approach, interference alignment involves the design of the appropriate
vectors for different users such that: i) the interfering
terms at each receiver are squeezed into a subspace of the available signal
space at that receiver, and ii) the interference subspace can be
separated from the desired signal subspace. Signal space alignment
is applicable to ICs with multiple antennas or ICs with time
varying/frequency selective channel coefficients. Signal scale
alignment, on the other hand, uses structured coding, e.g.,
lattice codes, to align interference at the signal level and is particularly useful for the case of single antenna constant IC (not varying with time/frequency).

For the fully connected $K$-user Gaussian IC ($K>2$), most of the effort has focused on
the characterization of the DoF. The DoF for a Gaussian
IC shows the growth of the maximum achievable sum rate in the limit
of increasing Signal to Noise Ratio ($\SNR$). In \cite{HostMadsen}, Host-Madsen
and Nosratinia showed that the DoF of the $K$-user Gaussian IC is less than or equal to $\frac{K}{2}$. They also conjectured that for the fully connected $K$-user \emph{constant} Gaussian IC, the DoF
is less than or equal to unity regardless of the number of users. In \cite{ManytoOne}, for the special cases of many-to-one and one-to-many Gaussian ICs, the authors have computed the capacity region within constant bits. In their achievability scheme for the many-to-one Gaussian IC, they introduced the signal scale interference alignment technique. In \cite{CadambeJafarShamai}, using the signal scale interference alignment, the authors reported a class of fully connected real constant $K$-user Gaussian ICs with DoF arbitrarily close to $\frac{K}{2}$. In \cite{cadambe2008iaa}, using the idea of
signal space interference alignment, Cadambe and Jafar showed that for a fully connected $K$-user
Gaussian IC with time varying or frequency selective channel
coefficients, the DoF is equal to $\frac{K}{2}$, i.e., each
user can enjoy half of its available DoF in spite of interfering
signals from other users. Etkin and
Ordentlich in \cite{Etkin-Ordentlich} used some results of
additive combinatorics to show that for a constant fully connected
real Gaussian IC, the DoF is very sensitive to the
rationality/irrationality of channel coefficients. They showed that
for a fully connected constant real Gaussian IC with rational
channel coefficients, the DoF is strictly less than
$\frac{K}{2}$. Moreover, they showed that for a class of measure
zero of channel coefficients, the DoF is equal to
$\frac{K}{2}$. Independently, Motahari \emph{et al.} showed in
\cite{abolfazl-isit} that for a three-user constant symmetric real
Gaussian IC with irrational channel coefficients, the DoF is
equal to $\frac{3}{2}$. However, their assumption regarding the
channel symmetry restricted its scope to a subset of
measure zero of all possible channel coefficients. For a constant
Gaussian IC with complex channel coefficients, Cadambe \emph{et al.}
in \cite{CadambeAsymetric} showed that the Host-Madsen and Nosratinia
conjecture is not true. By introducing asymmetric complex
signaling, they proved that the $K$-user complex Gaussian IC with
constant coefficients has at least $1.2$ DoF for almost
all values of channel coefficients. Recently, Motahari \emph{et al.}
settled the problem in general case by proposing a new type of signal
scale interference alignment that can achieve $\frac{K}{2}$ DoF for almost all $K$-user real Gaussian ICs with constant
coefficients \cite{abolfazl-real,abolfazl-final}. The essence of
this new method, called {\em real alignment}, is
to align discrete points along a real axis based on some number-theoretic properties of rational and irrational numbers
\cite{abolfazl-final}.

It is straightforward to extend the results of \cite{cadambe2008iaa,abolfazl-final}  to the $K$-user MIMO interference channel with the same number of antennas at all nodes. In fact, based on the results of \cite{cadambe2008iaa,abolfazl-final}, it is not
difficult to see that for a $K$-user $M\times M$ MIMO Gaussian IC,
the DoF is equal to $\frac{KM}{2}$ whether the channel is
constant or time varying/frequency selective. However, extending
this conclusion to the general $K$-user $M\times N$ MIMO Gaussian IC
is not straightforward. In \cite{Gou-Jafar}, by using signal space
alignment in conjunction with the channel extension in time, the authors obtained a lower-bound on the DoF
of the $K$-user $M\times N$ time varying/frequency selective Gaussian IC. They also provided an upper-bound on the DoF of this channel which is valid for
both time/frequency varying and constant channel coefficients. The lower and the upper-bound in \cite{Gou-Jafar} coincide when $\frac{\max(M,N)}{\min(M,N)}$ is an integer. Another related work is \cite{Suh-Tse}
in which Suh and Tse considered the problem of interference
alignment for cellular networks. Using a method called subspace
interference alignment, they showed that as
the number of users in each cell increases, their achievable DoF also
increases and approaches the interference free DoF.

In this paper, we extend the results of \cite{Gou-Jafar} in two
directions. First, we show that their results can be extended to
\emph{constant} channels by generalizing the method of
\cite{abolfazl-final} to the MIMO case. Second, we improve their
results by introducing a higher achievable DoF and a tighter
upper-bound.

This paper is organized as follows: In section \ref{sec system
model}, the system model is introduced. In section \ref{sec main
results}, the main results are presented, followed by some
discussions. In section \ref{sec outerbound}, we present a new
upper-bound on the DoF of a MIMO Gaussian IC. In section
\ref{sec interfernce alignment}, we demonstrate our achievability result for
a three-user $1\times 2$ MIMO Gaussian IC and then generalize it
to the $K$-user $M\times N$ MIMO Gaussian IC. We will conclude in
section \ref{sec conclusions}.

\textbf{Notation}:  $\mathds{N}$, $\xxZ^+$ and $\xxZ$ represent the set of naturals, positive integers
and integers, respectively. The transpose of a vector $\mathbf{V}$ is denoted by $\mathbf{V}^T$.  For a set $\mathcal{S}$ and a real number $a$, we
define the set $a\cdot\mathcal{S}$ as: $$a\cdot\mathcal{S}\define\{a.s:\, s\in \mathcal{S}\}.$$ For two
sets $\mathcal{S}_1$ and $\mathcal{S}_2$, the set theoretic difference is denoted
by $\mathcal{S}_1\setminus \mathcal{S}_2=\{s\in \mathcal{S}_1: s \notin \mathcal{S}_2\}$. The union of two
sets $\mathcal{S}_1$ and $\mathcal{S}_2$ will be denoted by $\mathcal{S}_1\bigcup \mathcal{S}_2$. For two
positive integers $x$ and $y$, $\gcd(x,y)$ denotes the greatest
common divisor of $x$ and $y$. In addition, we use the following
notations:
\begin{displaymath}
 \xxK=\{1,\cdots,K\},\quad
\xxN=\{1,\cdots,N\},\quad \xxM=\{1,\cdots,M\},\quad \xxL=\{1,\cdots,L\}.
\end{displaymath}
\section{System Model}\label{sec system model}
We consider a constant fully connected real $K$-user MIMO Gaussian
IC. This channel is used to model a communication
network with $K$ transmitter-receiver pairs. Each transmitter which
is equipped with $M$ antennas wishes to communicate with its
corresponding receiver, which is equipped with $N$ antennas. All
transmitters share a common bandwidth and want to have reliable
communication at maximum possible rates. The channel output at the
$k^{th}$ receiver is characterized by the following input-output
relationship:
\begin{equation}\label{Eq:InputOuput}
\mathbf{Y}^{[k]}(t)=\xH^{[k1]}\xX^{[1]}(t)+\xH^{[k2]}\xX^{[2]}(t)+\cdots+\xH^{[kK]}\xX^{[K]}(t)+\xZ^{[k]}(t),
\end{equation}
where $t$ is the time index, $k\in \xxK$ is the user index,
$\mathbf{Y}^{[k]}=(Y^{[k]}_1,\cdots,Y^{[k]}_N)^T$ is the $N \times 1$ output
signal vector of the $k^{th}$ receiver,
$\xX^{[j]}=(X^{[j]}_1,\cdots,X^{[j]}_M)^T$ is the $M \times 1$ input signal
vector of the $j^{th}$ transmitter, $\xH^{[kj]}=[h^{[kj]}_{nm}]$ is the
$N \times M$ channel matrix between transmitter $j$ and receiver $k$
with the $(n,m)^{th}$ entry specifying the channel gain from the $m^{th}$
antenna of transmitter $j$ to the $n^{th}$ antenna of receiver $k$, and
$\mathbf{Z}^{[k]}=(Z^{[k]}_1,\cdots,Z^{[k]}_N)^T$ is $N\times 1$ additive
white Gaussian noise (AWGN) vector at the $k^{th}$ receiver. We
assume all noise terms are i.i.d. zero mean unit
variance real Gaussian random variables. It is assumed that each transmitter is subject to a power constraint $P$.

For a MIMO Gaussian IC with a power constraint $P$ at each transmitter, a $K$-tuple of rates $\mathbf{R}(P)=(R_1(P),\cdots,R_K(P))$
is said to be achievable if the transmitters
can increase the cardinalities of their message sets as
$2^{nR_i(P)}$ with block length $n$ and the average probability of error for all transmitters can be made arbitrarily small when $n$ is
sufficiently large. The capacity region of the
$K$-user MIMO Gaussian IC is the set of all achievable $K$-tuples $\mathbf{R}(P)$ and is denoted by $\mathscr{C}(P)$. Our primary objective in this paper is to characterize the sum capacity of this channel as $P\rightarrow\infty$.

For an achievable rate tuple $\mathbf{R}(P)=(R_1(P),\cdots,R_K(P))$, the corresponding achievable sum DoF (or simply achievable DoF) is defined as:
\begin{align}\dofl\triangleq\lim_{P\rightarrow\infty}\frac{\sum_{k=1}^K{R_k}(P)}{0.5\log(P)}.
\end{align}
The DoF of the channel is defined as the supremum of all achievable DoF. More precisely,
\begin{align}\dof\define\lim_{P\rightarrow\infty}\sup_{\mathbf{R}(P)\in
\mathscr{C}(P)}\frac{\sum_{k=1}^K{R_k}(P)}{0.5\log(P)}.
\end{align}
In other words, $\dof$ represents the
maximum achievable sum rate as $\SNR$ goes to infinity. For notational consistency, an upper-bound on DoF will be denoted by $\dofu$.

In the sequel, a $(K,M\times N)$ IC refers to a constant fully connected $K$-user MIMO Gaussian IC with $M$ antennas at each transmitter and $N$ antennas at each receiver.
\section{Main Result and discussions}\label{sec main results}
The main results of the paper are formulated in the following two theorems:
\begin{theorem}\label{Thm:upper bound}
The DoF of a $(K,M\times N)$ IC is upper-bounded by:
\begin{align}\label{Eq:DoFUpper}
  \dofu\triangleq K\min\left\{\max(M,N)\rho^+,\min(M,N)(1-\rho^-)\right\},\end{align} where $\rho^+$ and $\rho^-$ are given by:
\begin{align}\label{Eq:Best_rho}
\begin{split}
\rho^-=\max_{n\in\xxK}\frac{\lfloor n\rho_0\rfloor}{n},\qquad
\rho^+=\min_{n\in\xxK}\frac{\lceil n\rho_0\rceil}{n},
\end{split}
\end{align}
and where $\rho_0\define\frac{\min(M,N)}{M+N}$ and  $\lfloor\cdot\rfloor$ and $\lceil\cdot\rceil$ are respectively the floor and the ceiling functions.
\end{theorem}
\begin{proof} : See section \ref{sec outerbound}.\end{proof}
\begin{theorem}\label{Thm:achievability theorem}
For a $(K,M\times N)$ IC, we can achieve $\dofl$ degrees of freedom
for \emph{almost all} channel realizations where:
\begin{align}\label{Eq:DoF Achiev}
    \dofl=\left\{
                      \begin{array}{ll}
                        K\min(M,N)\,\min(1,\frac{\beta}{K}), & \hbox{$K< \beta+1$} \\
                        K\,\frac{MN}{M+N}, & \hbox{$K\geq \beta+1$}
                      \end{array},
                    \right.
\end{align}
and where $\beta\define\frac{\max(M,N)}{\min(M,N)}$.
\end{theorem}
\begin{proof}
It is easy to show that in a $(K,M\times N)$ IC, one can always achieve $$\min\left\{\max(M,N), K\min(M,N)\right\}$$ DoF by zero-forcing. In section \ref{sec interfernce alignment}, we prove that using real interference alignment, we can almost surely achieve $K\frac{MN}{M+N}$ DoF for a $(K,M\times N)$ IC. By combining these two results, we obtain (\ref{Eq:DoF Achiev}).
\end{proof}
\begin{remark}\label{Remark1}
If $K\geq K_u\define\frac{M+N}{\gcd(M,N)}$, then we will have $\rho^-=\rho^+=\rho_0$ in Theorem \ref{Thm:upper bound} and consequently $\dofu= K\frac{MN}{M+N}$. On the other hand, since $K_u\geq \beta+1 $, from Theorem \ref{Thm:achievability
theorem}, we have $\dofl= K \frac{MN}{M+N}$ for $K\geq K_u$. Hence, for $K\geq K_u$, the channel DoF is equal to  $K\frac{MN}{M+N}$.\end{remark}

\begin{remark}\label{Remark2}
For $K\leq \beta+1$, one can easily verify that $\rho^-=0$ and $\rho^+=\frac{1}{K}$, and therefore, from (\ref{Eq:DoFUpper}), the DoF is upper-bounded by: $$\dofu= K\, \min\left\{\frac{\max(M,N)}{K},\min(M,N)\right\}=K\min(M,N)\,\min(1,\beta/K).$$ Combining with Theorem \ref{Thm:achievability theorem}, we see that for $K<\beta+1$, the DoF is equal to $K\min(M,N)\,\min(1,\beta/K)$. Let us define: $$K_l\define \lfloor\beta\rfloor+1.
$$ While our results provide a complete characterization of DoF for $K\geq K_u$ and
$K\leq K_l$, this characterization for the case of $K_l<K<K_u$
seems to be challenging. Our achievable DoF is not generally tight in this range.
\end{remark}

\begin{remark}
Consider the case that
$\beta$ is an integer. In this case,
$\gcd(M,N)~=~\min(M,N)$, and hence, $K_u=\beta+1$.
Therefore, according to Remark \ref{Remark1}, for $K\geq \beta+1$, the
DoF is equal to $K\,\frac{MN}{M+N}$.
On the other hand, since $\min(1,\beta/K)=1$ for $K\leq \beta$, it follows from Remark \ref{Remark2} that $\dof= K \min(M,N)$ for $K\leq \beta$. Hence, we have an exact characterization of DoF when $\beta$ is an integer.
\end{remark}

\begin{remark}
For $K<\beta+1$, the achievable scheme in Theorem \ref{Thm:achievability theorem} is merely based on zero-forcing and no interference alignment is required. For $K\geq \beta+1$, our achievable scheme is based on the real interference alignment \cite{abolfazl-final}.
\end{remark}
\section{upper-bound on the DoF for the $K$-user MIMO interference
channel}\label{sec outerbound}
In this section, we prove Theorem
\ref{Thm:upper bound} which provides a new upper-bound on the DoF of the $(K,M\times N)$ Gaussian IC. Our method is based on the averaging argument of \cite{HostMadsen} which is generalized to the MIMO case in \cite{Gou-Jafar}.

Consider a $(W,M\times N)$ Gaussian IC where $W\leq K$ is a constant. We divide these $W$ users into two disjoint sets of size $W_1$ and $W_2$, where $W=W_1+W_2$. Let us assume that the transmitters in each set are cooperating, and the receivers in each set are cooperating as well.
This results in a two-user
MIMO Gaussian IC with $W_1M$, $W_2M$ antennas
at transmitters and $W_1N$, $W_2N$ antennas at their corresponding receivers. It is
proved in \cite{Jafar-Fakhereddin} that for a two-user MIMO Gaussian IC with $M_1$, $M_2$ antennas at transmitter 1, 2 and $N_1$, $N_2$ antennas at their corresponding receivers, the DoF is equal to:
\begin{equation}\label{Eq:2user upperbound}
J(M_1,M_2,N_1,N_2)=\min\{M_1+M_2,N_1+N_2,\max(M_1,N_2),\max(M_2,N_1)\}.
\end{equation}
Since cooperation does not reduce the capacity, the DoF of the
original $W$-user interference channel does not exceed
$J(W_1M,W_2M,W_1N,W_2N)$. Thus, for any $i_1,i_2,\cdots,i_W\in\xxK$,
$i_1\neq i_2\neq\cdots\neq i_W$, we have:
\begin{equation}\label{Eq:L user upperbound}
    d_{i_1}+d_{i_2}+\cdots+d_{i_W}\leq J(W_1M,W_2M,W_1N,W_2N),
\end{equation}
where $d_k$ denotes the DoF of user $k$. Adding up all inequalities similar to (\ref{Eq:L user upperbound}),
the DoF of the $K$-user Gaussian IC is upper-bounded as:
\begin{equation}\label{Eq:K user upperbound}
    \dof\leq\frac{K}{W}J(W_1M,W_2M,W_1N,W_2N).
\end{equation}
It is proved in Appendix \ref{App:Upper-boundProof} that the function $J(W_1M,W_2M,W_1N,W_2N)$ can
be upper-bounded as:
\begin{equation}\label{Eq:2user upperbound2}
   J(W_1M,W_2M,W_1N,W_2N)\leq \max\{\max(M,N)W_{\text{min}},
   \min(M,N)W_{\text{max}}\},
\end{equation}
where $W_{\text{max}}=\max(W_1,W_2)$ and $W_{\text{min}}=\min(W_1,W_2)$. Combining (\ref{Eq:2user upperbound2}) and (\ref{Eq:K user upperbound}), we have:
\begin{equation}\label{Eq:K user upperbound2}
   \dof\leq K\,G(\rho),
\end{equation}
\begin{figure}
  \begin{center}
  \includegraphics[width=10cm]{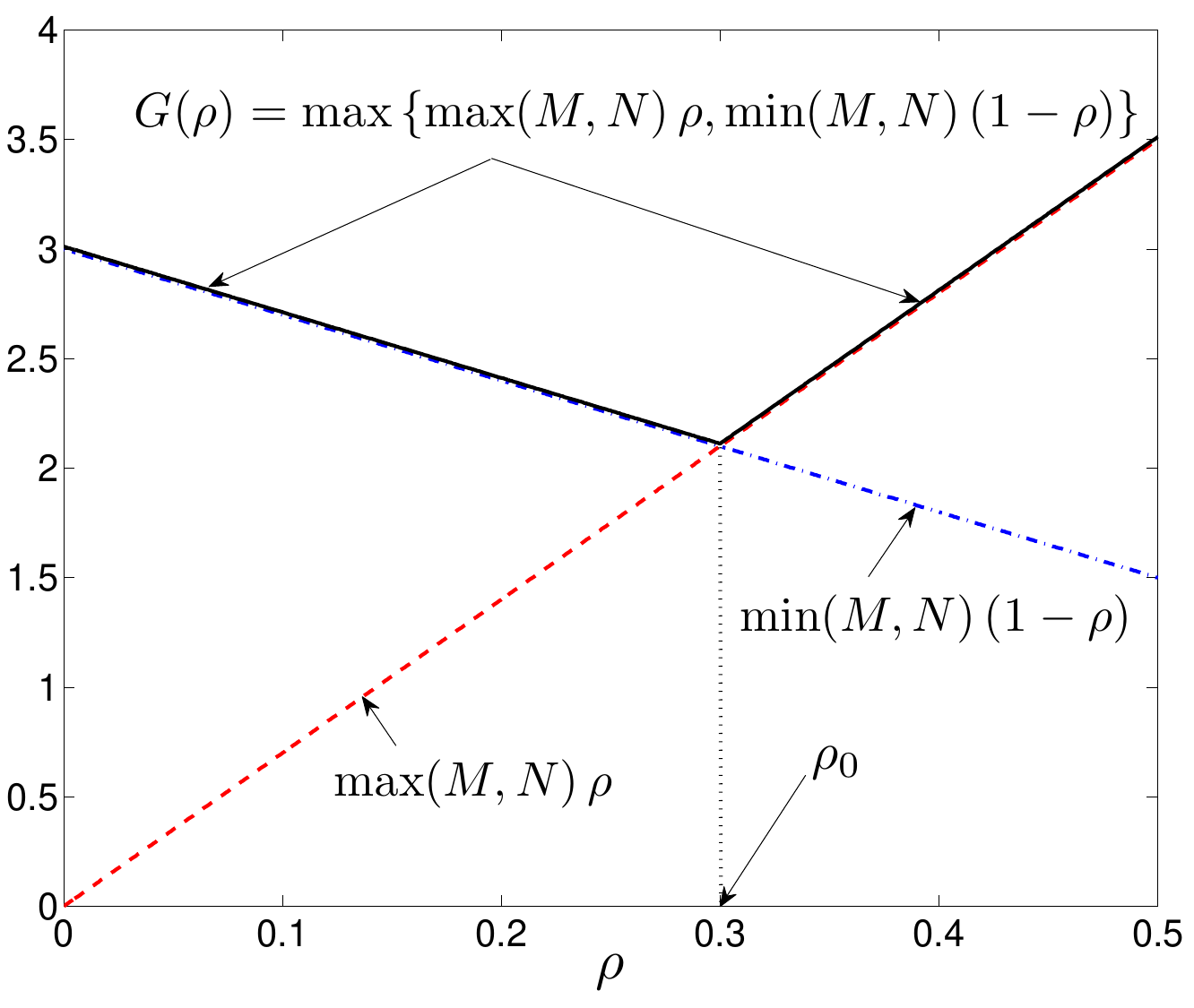}\\
  \caption{Typical shape of function $G(\rho)$ in (\ref{Eq:G(ru)}).}\label{Fig:typical shape of upper}
  \end{center}
\end{figure}
where $\rho\define\frac{W_{\text{min}}}{W}$ and
\begin{align}\label{Eq:G(ru)}
G(\rho)\define\max\{\max(M,N)\rho,
   \min(M,N)(1-\rho)\}.
 \end{align}
 A typical plot of $G(\rho)$ is depicted in Fig. 1. To obtain the tightest upper-bound, we need to minimize $G(\rho)$ over the rational number $\rho$. However, there are two constraints on $\rho$: \\
 $\quad$C1) $0\leq\rho\leq\frac{1}{2}$,\\
 $\quad$C2) the denominator of $\rho$ as a rational number in lowest terms can not exceed $K$. 
 
Thus, the goal is to minimize $G(\rho)$ subject to the constraints C1 and C2. It is straightforward to show that (see also Fig.\ref{Fig:typical shape of upper}) without any constraint on $\rho$, the function $G(\rho)$ is minimized when:
\begin{equation}
\max(M,N)\rho=\min(M,N)(1-\rho).
\end{equation}
Equivalently, $G(\rho)$ is minimized at $\rho=\rho_0$, where $\rho_0$ was defined in Theorem \ref{Thm:upper bound}.  Although $\rho=\rho_0$ satisfies constraint C1, it does not generally satisfy constraint C2 because the denominator of $\rho_0$ in the simplest form can exceed $K$. Therefore, to find the optimal $\rho$ that minimizes $G(\rho)$ subject to the constraints C1 and C2, we need to find the closest rational neighbors of $\rho_0$ with denominator not exceeding $K$. Let $\rho^-$ and $\rho^+$ denote the closest rational neighbors of $\rho_0$ with denominator not exceeding $K$ such that $0\leq\rho^-\leq \rho\leq \rho^+$. From (\ref{Eq:K user upperbound2}), for such $\rho^+$ and $\rho^-$, we have:
  \begin{align}
  \begin{split}
\dof&\leq K \max\{\max(M,N)\rho^+,
   \min(M,N)(1-\rho^+)\}=K\max(M,N)\rho^+\\
   \dof&\leq K \max\{\max(M,N)\rho^-,
   \min(M,N)(1-\rho^-)\}=K\min(M,N)(1-\rho^-)
   \end{split} .
\end{align}
Therefore, the final upper-bound can be expressed as:
\begin{align}\label{Eq:final upper}
\dof\leq K\min\left\{\max(M,N)\rho^+,\min(M,N)(1-\rho^-)\right\}.
\end{align}
The problem of finding the closest rational neighbors of a real number with denominator less than or equal to $K$ is addressed in the following lemma:
 \begin{lemma}\label{Thm:RationalApprox}
Let $\alpha\in(0,1)$ be a real number. Given a positive integer $K$, the closest rational neighbors of $\alpha$ ($\alpha^-\leq\alpha\leq \alpha^+$) with denominator not exceeding $K$ are given by:
\begin{align}
 \alpha^-&=\max_{n\in \{1,2,\cdots,K\}}\frac{\lfloor n\alpha \rfloor}{n}\label{eq:neigbours1},\\
 \alpha^+&=\min_{n\in\{1,2,\cdots,K\}}\frac{\lceil  n\alpha \rceil}{n}\label{eq:neigbours2}.
\end{align}
\end{lemma}
\begin{proof} See Appendix \ref{App:RationalApprox}. \end{proof}
Now, (\ref{Eq:Best_rho}) easily follows from the above lemma and the proof is complete.

\begin{figure}
\centering
\subfigure[$K=5$]{
\includegraphics[scale=.33]{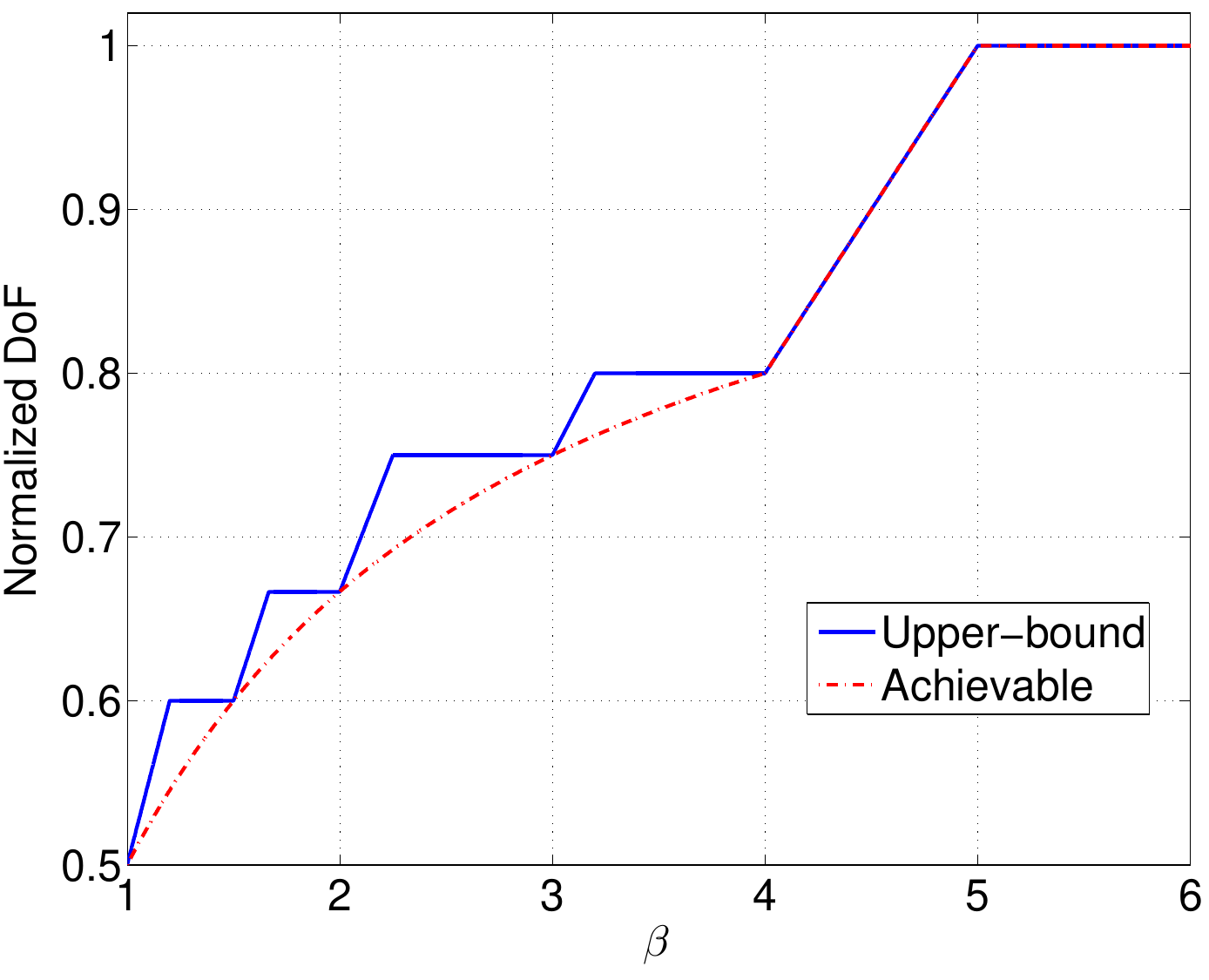}
}
\subfigure[$K=10$]{
\includegraphics[scale=.33]{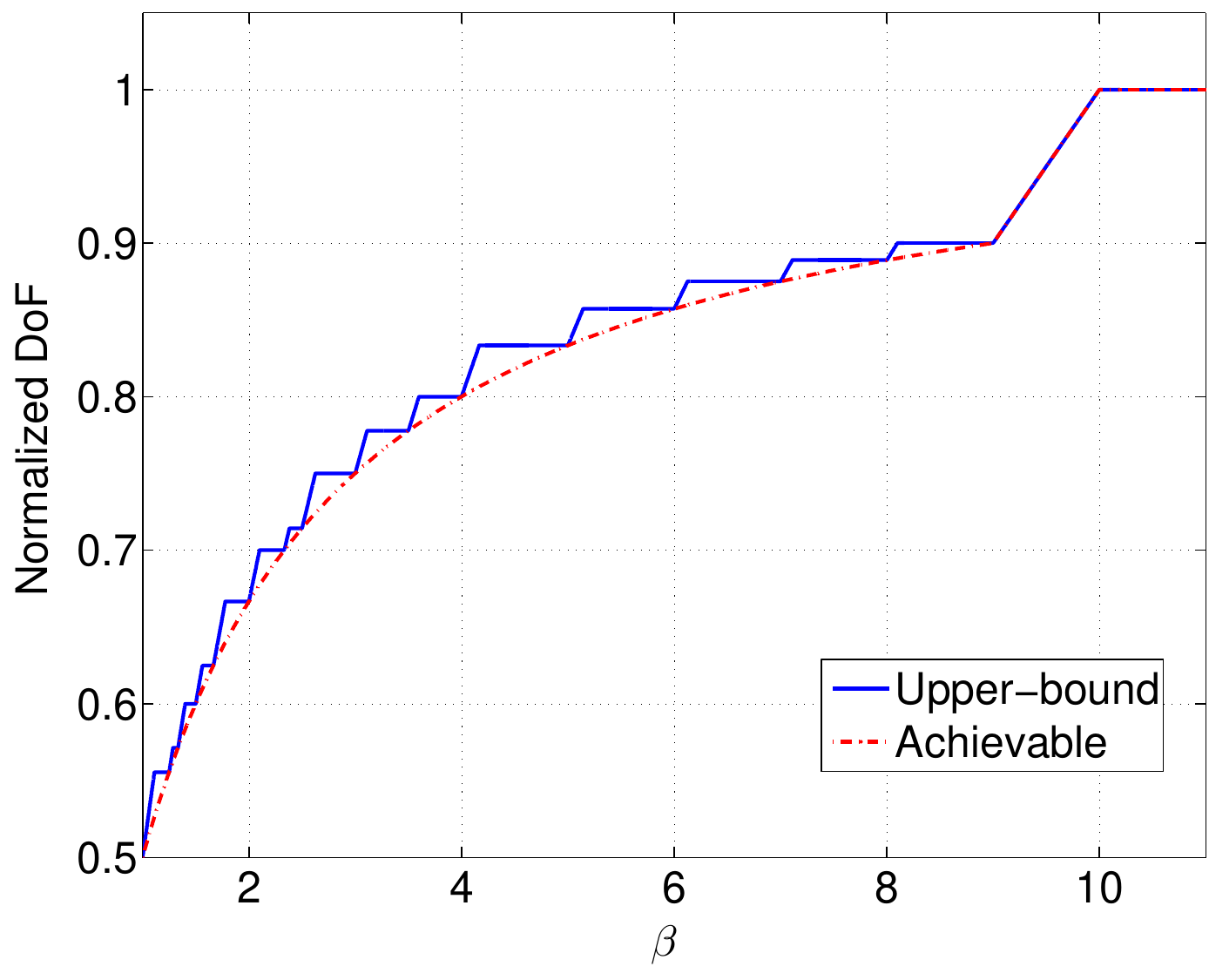}
}
\caption{Our achievable and upper-bound on normalized DoF of a $(K,M\times N)$ IC for $K=5$ and $K=10$.}\label{Fig:Normalized DoF}
\end{figure}

The upper-bound in (\ref{Eq:final upper}) can be pictorially presented in a more elegant way by defining the \emph{normalized degrees of freedom}. The normalized DoF of a $(K,M\times N)$ IC is defined as:
\begin{align}\dof_{\text{norm}}\define\frac{\dof}{K\min(M,N)}.\end{align}
Note that $K\min(M,N)$ is the DoF of a system consisting of $K$ non-interfering $M\times N$ MIMO channels.
Therefore, $\dof_{\text{norm}}$ is always less than unity. Unlike $\dofu$ which is a function of three parameters $M,N,$ and $K$, the normalized upper-bound $\dofu_{\text{norm}}$ is a function of only two parameters $K$ and $\beta$. Specifically, from (\ref{Eq:DoFUpper}), we have:
\begin{align}\label{Eq:normalized DoF}
    \dofu_{\text{norm}}= \min\{\beta\rho^+,1-\rho^-\},
\end{align}
where $\rho^-$ and $\rho^+$ are obtained from (\ref{Eq:Best_rho}) with $\rho_0=\frac{1}{\beta+1}$. According to  Theorem \ref{Thm:achievability theorem}, our achievable normalized DoF can be expressed as:
\begin{align}\label{Eq:NDoF Achiev}
    \dofl_{\text{norm}}=\left\{
                      \begin{array}{ll}
                        \min(1,\frac{\beta}{K}), & \hbox{$K< \beta+1$} \\
                        \frac{\beta}{\beta+1}, & \hbox{$K\geq \beta+1$}
                      \end{array}.
                    \right.
\end{align}
Two examples comparing our achievable result and upper-bound on $\dof_{\text{\text{norm}}}$ are depicted in Fig. \ref{Fig:Normalized DoF}.
\section{Achievability Scheme for Theorem \ref{Thm:achievability theorem}}\label{sec interfernce alignment}
In this section, we prove Theorem \ref{Thm:achievability theorem} and
examine the interference alignment method that achieves
$\frac{MN}{M+N}K$ DoF for almost all channel
realizations. To explain the key ideas, we start with the simple example of
a $(3,1\times 2)$ system.

A new method for interference alignment has been recently introduced
by Motahari \emph{et al.} in \cite{abolfazl-final}.
By applying arguments from the field of Diophantine approximation in
Number Theory, they showed that interference alignment can be
performed based on the properties of rational and irrational
numbers. Using this new type of alignment, which the authors called \emph{real interference alignment}, the DoF of the $K$-user \emph{constant} Gaussian IC with
single antenna can be achieved for almost all channel realizations. Since our achievability scheme is based on an extension of real interference alignment, we first review the basic ideas behind this technique. In our discussions, we follow the footsteps of \cite{abolfazl-final} and \cite{Etkin-Ordentlich}.
\subsection{Preliminaries on Real Interference Alignment}
Real interference alignment essentially mimics, in one dimension, the basic rules of signal-space interference alignment. In signal space interference alignment, the transmit
signal of each user is a linear combination of some constant
vectors in Euclidean space, which hereafter will be called \emph{transmit directions},
where data determines the coefficients of this linear combination.
In this setup, interference alignment is realized by simultaneous design of appropriate
transmit directions for different users such that:\\
i) Interfering signals from other users are received aligned at the intended receiver. In other words, all interfering terms at each receiver fall into a subspace of the available signal
space at that receiver. This condition will be referred to as \emph{alignment condition}.\\
ii) The interference subspace can be separated from the desired signal subspace at each receiver.
 This condition will be referred to as \emph{separability condition}.

  Note that transmit directions are selected according to the channel coefficients. In signal space alignment, when both alignment and separability conditions are satisfied, we can separate the desired signal from  interfering signals by zero-forcing. This is achieved by projecting the received signal onto the subspace which is orthogonal to the interference subspace.

    Consider a $K$-user Gaussian IC with a single antenna at all nodes where channel coefficients are all constant. 
     Since each node relies on a one-dimensional signal space, we are essentially dealing with real numbers instead of vectors and alignment should happen at the \emph{signal level}. Recall that the $n$-dimensional Euclidean space is a vector space over the field of real numbers. We can similarly consider the field of real numbers as a vector space over the field of rational numbers.
To introduce the counterparts of separability and alignment conditions in real interference alignment, we need the notion of \emph{rationally independence}.
\begin{definition}[rationally independence]
The real numbers $\omega_1,\omega_2,\cdots,\omega_m$ are said to be rationally independent if whenever integers $k_1, k_2, ... , k_m$ satisfy $$k_1 \omega_1 + k_2 \omega_2 + \cdots + k_m \omega_m = 0,$$ we should have $k_i = 0$ for $i = 1,\cdots, m$, i.e., the only representation of  zero as a linear combination of $\omega_i\,,i=1,\cdots,m$ is the trivial solution.
\end{definition}
If a given set of real numbers $\omega_1,\omega_2,\cdots,\omega_m$ are not rationally independent, they can be represented as rational linear combinations of a minimum number, say $n$, of some fixed rationally independent real numbers ($n<m$). Here $n$ is called the \emph{rational dimension} of real numbers $\omega_i,\,i=1,\cdots,m$. The notion of rational dimension is defined precisely in the following.
\begin{definition}[rational dimension]
The rational dimension of real numbers $\omega_1,\omega_2,\cdots,\omega_m$ is defined as the smallest natural number $n$ such that all numbers $\omega_i,\,i=1,\cdots,m$ can be represented as  rational linear combinations of $n$ fixed rationally independent real numbers. The rational dimension of a set $\sA$ of real numbers will be denoted by $\dim(\sA)$.
\end{definition}
Suppose that $\omega_1,\omega_2,\cdots,\omega_m$ are rationally independent real numbers. Therefore, for arbitrary integers $k_1, k_2, ... , k_m$, not all of them equal to zero, we have $|k_1 \omega_1 + k_2 \omega_2 + \cdots + k_m \omega_m|>0.$ The problem of finding a non-zero lower-bound on the absolute value of an integer linear combination of rationally independent real numbers is closely related to metric Diophantine approximation in Number Theory \cite{Bernik}. The following theorem which is a special case of Khintchine-Groshev Theorem in metric Diophantine approximation \cite{Bernik} provides a quantitative lower-bound on the absolute value of a linear combination of real numbers.
\begin{theorem}[Khintchine-Groshev]
Assume $\epsilon>0$ is an arbitrary positive constant. For \emph{almost all} $l$-tuples $\boldsymbol{\omega} = (\omega_1, \omega_2,\cdots,\omega_l)$ of real numbers, one can find a constant $c$ such that the
inequality
\begin{align}\label{Eq:Khintchine-Groshev}|p+q_1 \omega_1 + q_2 \omega_2 + \cdots + q_l \omega_l|>\frac{c}{
(\max_i q_i )^l}\end{align}
holds for all  $p\in\mathds{Z}$ and all $q = (q_1, q_2,\cdots,q_l)\in \mathds{Z}^l\setminus\mathbf{0}$.
\end{theorem}
It is important to note that the Khintchine-Groshev Theorem is valid for ``almost all" real numbers. That is the Lebesgue measure of those real numbers satisfying the Khintchine-Groshev Theorem is one. It should be pointed out here that the Khintchine-Groshev Theorem is not valid even for all rationally independent real numbers.

The real numbers $\omega_i,\,i=1,\cdots,l$, in the Khintchine-Groshev Theorem could be independent quantities or they can lie on some well-behaved manifold. Specifically, the Khintchine-Groshev Theorem is valid when all the real numbers $\omega_i,\,i=1,\cdots,l$ are different monomials in $m<l$ independent variables \cite{abolfazl-final}\cite{Beresnevich}.

Consider two sets $\sA$ and $\sB$ of real numbers with rational dimensions $\dim(\sA)$ and $\dim(\sB)$, respectively. We define the \emph{alignment index} of $\sA$ and $\sB$, which is denoted by  $\chi(\sA,\sB)$, as: $$\chi(\sA,\sB)\define\frac{\dim(\sA\bigcup \sB)}{\max(\dim(\sA),\dim(\sB))}.$$ It is easy to see that  $\chi(\sA,\sA)=1$ for any non-empty set $\sA$. Furthermore, one can readily see that $\chi(\sA,\sB)\geq 1$ for any two non-empty sets $\sA$ and $\sB$. The alignment index of more than two sets is similarly defined as the ratio of the rational dimension of their union to the maximum of the individual rational dimensions.

Now, consider two sequences $\sA_n$ and $\sB_n$ of sets where the cardinalities of $\sA_n$ and $\sB_n$ grows to infinity as $n\rightarrow\infty$. We define the notion of \emph{asymptotic alignment} as follows:
\begin{definition}[Asymptotic alignment]
Two sequences $\sA_n$ and $\sB_n$ of sets are called asymptotically aligned if $\limsup_{n\rightarrow\infty}\chi(\sA_n,\sB_n)=1$.
\end{definition}
The above definition can be generalized to more than two sequences of sets. In other words, $S$ sequences of sets $\sA^{[1]}_n,\cdots,\sA^{[S]}_n$ are call asymptotically aligned if the $\limsup$ of their alignment index goes to unity as $n\rightarrow\infty$.

Consider two sequences of discrete random variables $X_n$ and $Y_n$ that are uniformly distributed over $\sA_n$ and $\sB_n$, respectively. If $\sA_n$ and $\sB_n$ are asymptotically aligned, the random sequences $X_n$ and $Y_n$ will be called asymptotically aligned.

\begin{example} Consider the following sequences of sets: $$\sA_n=\left\{a_1^{n_1}a_2^{n_2}a_3^{n_3}:\,\,n_i\in\{0,1,\cdots,n\}\right\},\quad n=1,2,\cdots$$
where $a_1,a_2,$ and $a_3$ are selected as three rationally independent real numbers such that for every $n$ all the elements of $\sA_n$ are rationally independent. According to the Khintchine-Groshev theorem, almost all triples of real numbers satisfy this condition. One can easily confirm that $\dim(\sA_n)=(n+1)^3$. Under this condition, the two sequences $a_1\cdot\sA_n$ and $a_2\cdot\sA_n$ of sets are asymptotically aligned. The reason is that $\left[a_1\cdot\sA_n\bigcup a_2\cdot\sA_n\right]\subset A_{n+1}$ and hence $\chi(a_1\cdot\sA_n,a_2\cdot\sA_n)\leq\frac{(n+2)^3}{(n+1)^3}$ which tends to one as $n\rightarrow\infty$.\end{example}
\subsection{Sketch of Proof for a $(3,1\times 2)$ System}
\begin{figure}
 \centering
 \includegraphics[width=6.5cm]{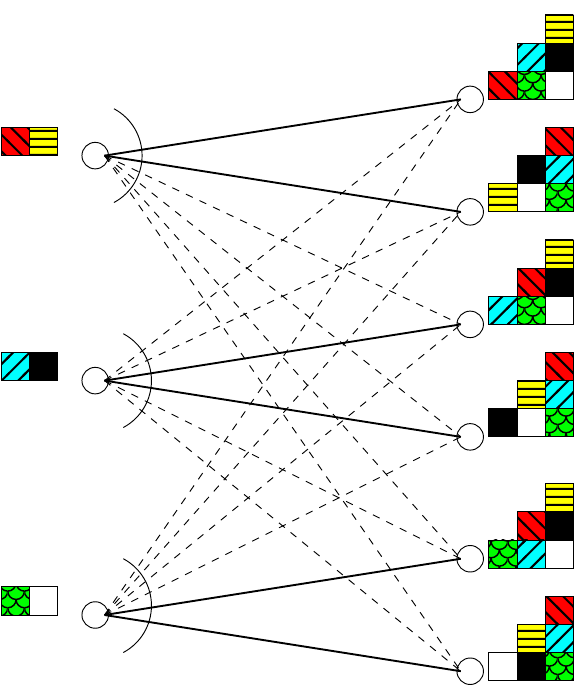}\\
  \caption{Real Interference alignment for a $(3,1\times 2)$
 Gaussian IC: the transmit signal of each user is composed of two independent
 parts which are depicted here by two adjacent squares.
 By the real interference alignment, the squares in each column at the receiver side are approximately aligned.}
\label{fig1}
\end{figure}

In this part, we explain our achievability scheme for a $(3,1\times 2)$ system. This system is depicted in Fig.~\ref{fig1}. The rigorous proof of our achievability scheme will be provided in the next part.\\
The transmit signal of each user is a weighted sum of two
independent parts: the first part is intended for the first receive
antenna and the second part is intended for the second receive
antenna. The weights are corresponding channel
coefficients. That is the transmit signal of user $k$ can be
expressed as:
\begin{align}\label{3user-transmit}
    X^{[k]}&=h^{[kk]}_{11}X^{[k]}_{1}+h^{[kk]}_{21}X^{[k]}_{2},\,\,k=1,2,3.
\end{align}
As we shall see later in more details, the transmission scheme is such that the following conditions are satisfied (see Fig. \ref{fig1}):
\begin{itemize}
  \item At the first receive antenna of user-$1$:
\begin{itemize}
  \item signals $h^{[22]}_{11}X^{[2]}_{1}$
and $h^{[33]}_{11}X^{[3]}_{1}$ are received$^{\ast}$ asymptotically aligned, and
  \item signals $h^{[11]}_{21}X^{[1]}_{2},h^{[22]}_{21}X^{[2]}_{2}$ and $h^{[33]}_{21}X^{[3]}_{2}$ are received\symbolfootnote[1]{after multiplication with the corresponding channel coefficients.} asymptotically aligned.
\end{itemize}
  \item At the second receive antenna of user-$1$:
  \begin{itemize}
  \item signals $h^{[22]}_{21}X^{[2]}_{2}$
and $h^{[33]}_{21}X^{[3]}_{2}$ are received$^{\ast}$ asymptotically aligned, and
  \item signals $h^{[11]}_{11}X^{[1]}_{1},h^{[22]}_{11}X^{[2]}_{1}$ and $h^{[33]}_{11}X^{[3]}_{1}$ are received\symbolfootnote[1]{after multiplication with the corresponding channel coefficients.} asymptotically aligned.
\end{itemize}
\end{itemize}
It is obvious that a similar statement is valid for the other users.
At the first receive antenna of user-1, we have the sum of following terms:
\begin{itemize}
 \item the contribution of $h^{[11]}_{11}X^{[1]}_{1}$,
 \item the aligned contribution of $\{h^{[22]}_{11}X^{[2]}_{1},h^{[33]}_{11}X^{[3]}_{1}\}$, and
 \item the aligned contribution of
$\{h^{[11]}_{21}X^{[1]}_{2},h^{[22]}_{21}X^{[2]}_{2},h^{[33]}_{21}X^{[3]}_{2}\}$.
\end{itemize}

Provided that these three parts can be successfully decoded, each of
them occupies almost $\frac{1}{3}$ of the available DoF\symbolfootnote[2]{Note that the available DoF at each receiver is equal to $1$.} at the first receive
antenna of user-1.
Therefore, the desired part, namely
$h^{[11]}_{11}X^{[1]}_{1}$, has a share of almost $\frac{1}{3}$ of the available DoF.
Similarly, at the second receive antenna of user-1, the desired signal $h^{[11]}_{21}X^{[1]}_{2}$ has a share of almost $\frac{1}{3}$ of the available DoF. Hence, we can
achieve the DoF of $\frac{2}{3}$ per user.\\
To align the signals as described above, we need to further divide each signal $X^{[k]}_i,\,i=1,2$ into several components. Further details will be provided in the following.
\subsection{Proof of Theorem \ref{Thm:achievability theorem}}
Consider a $(K,M\times N)$ IC where each user satisfies a power constraint $P$. For any $\epsilon>0$, we will provide a transmission scheme that achieves $\sum_{k=1}^{K}R_k=\frac{KMN}{M+N}(\frac{1}{2}-\epsilon)\log_2 P-o(\log_2 P)$, showing that $\dof\geq \frac{KMN}{M+N}$.

In our achievable scheme, each transmitter uses its antennas separately, i.e., there is no cooperation among transmit antennas of each user.
In fact, user $k$ relies on $M$ independent codebooks $\mathcal{C}^{[k]}_{m}(P,\epsilon,\tau),\,m=1,\cdots,M$, of block length $\tau$ where $\mathcal{C}^{[k]}_{m}(P,\epsilon,\tau)$ is associated with its $m^{\text{th}}$ transmit antenna. Each codebook $\mathcal{C}^{[k]}_{m}(P,\epsilon,\tau),\,m\in\xxM$, is obtained by a linear combination of  $N$ independent sub-codebooks  $\mathcal{C}^{[k]}_{mn}(P,\epsilon,\tau),\,n=1,\cdots,N$. More precisely, the transmit symbol from the $m^{\text{th}}$ antenna of user $k$ at time index $t$ can be expressed as:
\begin{align}\label{Eq:Xm}
X^{[k]}_{m}(t)=\sum_{n=1}^{N}h^{[kk]}_{nm}X^{[k]}_{mn}(t),\quad t=1,\cdots,\tau,
\end{align}
where $\left(X^{[k]}_{m}(1),\cdots,X^{[k]}_{m}(\tau)\right)\in\mathcal{C}^{[k]}_{m}(P,\epsilon,\tau)$ and $\left(X^{[k]}_{mn}(1),\cdots,X^{[k]}_{mn}(\tau)\right)\in\mathcal{C}^{[k]}_{mn}(P,\epsilon,\tau)$. The sub-codebook $\mathcal{C}^{[k]}_{mn}(P,\epsilon,\tau)$ is intended to be decoded at the $n^{\text{th}}$ receive antenna of user $k$. Each sub-codebook $\mathcal{C}^{[k]}_{mn}(P,\tau)$ is in turn obtained by adding  $L$ independent sub-sub-codebooks $\mathcal{C}^{[k]}_{mnl}(P,\epsilon,\tau),\,l=1,\cdots,L,$ i.e.,
\begin{align}\label{Eq:Xmn}
X^{[k]}_{mn}(t)=\sum_{l=1}^{L}X^{[k]}_{mnl}(t),\quad t=1,\cdots,\tau,
\end{align}
where $\left(X^{[k]}_{mnl}(1),\cdots,X^{[k]}_{mnl}(\tau)\right)\in\mathcal{C}^{[k]}_{mnl}(P,\epsilon,\tau)$ and $L\in\mathds{N}$ is a design parameter which will be determined later.
Each sub-sub-codebook $\mathcal{C}^{[k]}_{mnl}(P,\epsilon,\tau)$ is generated i.i.d. according to a uniform distribution over $\Lambda^{[k]}_{mnl}(P,\epsilon)$, where:
\begin{align}\label{Eq:LambdaSet}
   \Lambda^{[k]}_{mnl}(P,\epsilon)\define\gamma P^{\frac{
\nu-1+2\epsilon}{2(\nu+\epsilon)}}\omega^{[k]}_{mnl}\cdot\left\{-Q,-Q+1,\cdots,Q\right\},
\end{align}
in which:
\begin{itemize}
\item $Q\define\lfloor P^{\frac{1-\epsilon}{2(\nu+\epsilon)}}\rfloor$.
\item $\gamma$ is a normalizing constant selected such that the average transmit power of each user does not exceed $P$. In Appendix \ref{App:PowerConstraint}, we calculate the normalizing constant  $\gamma$ and show that it is independent of $\nu$ and $P$.
\item  $\nu\in\mathds{N}$ is an important design parameter which controls the cardinality of $\Lambda^{[k]}_{mnl}(P,\epsilon)$ as well as the magnitude of its elements. Since $|\Lambda^{[k]}_{mnl}(P,\epsilon)|=2Q+1\leq 2P^{\frac{1-\epsilon}{2(\nu+\epsilon)}}+1$, we refer to $\nu$ as the \emph{rate control parameter}.
\item $\omega^{[k]}_{mnl}$ is a real number which should be properly selected according to the channel coefficients for the purpose of interference alignment.
\end{itemize}

 Since $\gamma P^{\frac{
\nu-1+2\epsilon}{2(\nu+\epsilon)}}$ does not depend on $m, n,$ and $l$, the symbol $X^{[k]}_{mn}(t)$ can be considered as a random integer linear combination of $L$ real numbers $\omega^{[k]}_{mn1},\cdots,\omega^{[k]}_{mnL}$ multiplied by $\gamma P^{\frac{
\nu-1+2\epsilon}{2(\nu+\epsilon)}}$, i.e.,
\begin{align}\label{Eq:Xmn2}
X^{[k]}_{mn}(t)=\gamma P^{\frac{
\nu-1+2\epsilon}{2(\nu+\epsilon)}}\sum_{l=1}^{L}B^{[k]}_{mnl}\omega^{[k]}_{mnl},
\end{align}
where $B^{[k]}_{mnl}$'s are independently and uniformly distributed over $\{-Q,-Q+1,\cdots,Q\}$. Each $B^{[k]}_{mnl}$ will be referred to as \emph{a data stream}.
By substituting (\ref{Eq:Xmn2}) in (\ref{Eq:Xm}), the transmit symbol of user $k$ on its $m^{\text{th}}$ antenna can be reformulated as:
\begin{align}
	X^{[k]}_{m}(t)=\gamma P^{\frac{
\nu-1+2\epsilon}{2(\nu+\epsilon)}}\sum_{n=1}^{N}\sum_{l=1}^{L}B^{[k]}_{mnl}h^{[kk]}_{nm}\omega^{[k]}_{mnl}.
\end{align}
 We observe that $X^{[k]}_{m}(t)$ is a random integer linear combination of $NL$ real numbers $h^{[kk]}_{nm}\omega^{[k]}_{mnl}$, $n\in\xxN,\,l\in\xxL$. The real numbers $h^{[kk]}_{nm}\omega^{[k]}_{mnl},\,k\in\xxK,\,m\in\xxM,\,n\in\xxN,\,l\in\xxL$ act like beamforming vectors in signal space alignment and will be referred to as \emph{modulation pseudo-vectors}. Let us define $\Omega^{[k]}_{mn}$ as:
 \begin{align}
 \Omega^{[k]}_{mn}\define\left\{\omega^{[k]}_{mn1},\cdots,\omega^{[k]}_{mnL}\right\}.
 \end{align}
Since the $NL$ pseudo-vectors $h^{[kk]}_{nm}\cdot\Omega^{[k]}_{mn}$, $n\in\xxN$ carry independent data streams, they are required to be rationally independent, i.e.,
\begin{align}\dim\left(\bigcup_{n=1}^{N}\left[h^{[kk]}_{nm}\cdot\Omega^{[k]}_{mn}\right]\right)=NL,\quad \forall k\in\xxK\text{ and }\forall m\in\xxM.\end{align}
Using the above signaling scheme, the received signal at the $n^{\text{th}}$ antenna of receiver $k$ at time index $t$ can be expressed as:
\begin{align}
Y^{[k]}_n(t)&=\sum_{k'=1}^{K}\sum_{m=1}^{M}h^{[kk']}_{nm}X^{[k']}_{m}+Z^{[k]}_n(t)
=\gamma P^{\frac{
\nu-1+2\epsilon}{2(\nu+\epsilon)}}\sum_{k'=1}^{K}\sum_{m=1}^{M}\sum_{n'=1}^{N}\sum_{l=1}^L B^{[k']}_{mn'l}h^{[kk']}_{nm}h^{[k'k']}_{n'm}\omega^{[k']}_{mn'l}+Z^{[k]}_n(t)\label{Eq:Ykn1}\\
\begin{split}\label{Eq:Ykn2}
&=\gamma P^{\frac{
\nu-1+2\epsilon}{2(\nu+\epsilon)}}\Bigg[\underbrace{\sum_{m=1}^{M}\sum_{l=1}^{L}B^{[k]}_{mnl}\,(h^{[kk]}_{nm})^2\,\omega^{[k]}_{mnl}}_{\text{desired}}+\underbrace{\sum_{m=1}^{M}\sum_{\substack{
        n'=1\\  n'\neq n
      }}^{N}\sum_{l=1}^{L}B^{[k]}_{mn'l}\,h^{[kk]}_{nm}h^{[kk]}_{n'm}\,\omega^{[k]}_{mn'l}}_{\text{self-interference}}\\&\qquad\qquad\quad+\underbrace{\sum_{\substack{
        k'=1\\  k'\neq k
      }
}^{K}\sum_{m=1}^{M}\sum_{n'=1}^{N}\sum_{l=1}^L B^{[k']}_{mn'l}\,h^{[kk']}_{nm}h^{[k'k']}_{n'm}\,\omega^{[k']}_{mn'l}}_{\text{multi-user interference}}\Bigg]+Z^{[k]}_n(t).
\end{split}
\end{align}
As we see from (\ref{Eq:Ykn1}), the modulation pseudo-vectors from different transmit antennas of different users appear in $Y^
{[k]}_{n}(t)$ after multiplication with the corresponding channel coefficients. For example, the modulation pseudo-vector $h^{[k'k']}_{n'm}\omega^{[k']}_{mn'l}$ which is originated from the $m^{\text{th}}$ antenna of user $k'$ appears in $Y^{[k]}_{n}$ as $h^{[kk']}_{nm}h^{[k'k']}_{n'm}\omega^{[k']}_{mn'l}$.
We refer to $h^{[kk']}_{nm}h^{[k'k']}_{n'm}\omega^{[k']}_{mn'l}$ as a \emph{received pseudo-vector} in ${Y}^{[k]}_{n}(t)$. According to this terminology, $Y^{[k]}_{n}(t)$ is a noisy version of an integer linear combination of $LMNK$ received pseudo-vectors. Each received pseudo-vector has a data stream as its coefficient. We observe from (\ref{Eq:Ykn2}) that three different components appear in $Y^{[k]}_n(t)$:
\begin{itemize}
	\item The desired component which contains $LM$ data streams. Each desired data stream in $Y^{[k]}_n(t)$ (i.e., $B^{[k]}_{mnl}$) can be represented by an ordered pair $(m,l)$, $m\in\xxM$, $l\in\xxL$.
	\item The self-interference component which contains $LM(N-1)$ data streams. All data streams in this component are originated from transmitter $k$.
	\item The multi-user interference component which contains $LMN(K-1)$ data streams. All the data streams in this component are originated from interfering users.
\end{itemize}
 Let us define $\tilde{Y}^{[k]}_{n}(t)$ as the noise-free part of $Y^{[k]}_{n}(t)$. The received pseudo-vectors in $\tilde{Y}^{[k]}_{n}(t)$ are not necessarily rationally independent and therefore some of them may be expressed as rational linear combinations of the rest. Let us momentarily assume that $\tilde{Y}^{[k]}_{n}(t)$ is known at the $n^{\text{th}}$ antenna of receiver $k$. We then can recover a data stream from $\tilde{Y}^{[k]}_{n}(t)$ provided that its corresponding received pseudo-vector can not be represented as a rational linear combination of the other received pseudo-vectors in $\tilde{Y}^{[k]}_{n}(t)$. Accordingly, all the desired data streams at the $n^{\text{th}}$ antenna of receiver $k$ can be obtained from $\tilde{Y}^{[k]}_{n}(t)$ if the received pseudo-vectors $(h^{[kk]}_{nm})^2\omega^{[k]}_{mnl},\,m\in\xxM,\,l\in\xxL$ can not be expressed as rational linear combinations of $h^{[kk']}_{nm}h^{[k'k']}_{n'm}\omega^{[k']}_{mn'l},\,k'\in\xxK,\,m\in\xxM,\,n'\in\xxN,\,l\in\xxL,\,(k',n')\neq(k,n)$. This condition will be referred to as \emph{the separability condition for the $n^{\text{th}}$ antenna of receiver $k$}, parallel to the separability condition for signal space alignment. According to this terminology, if the separability condition holds at the $n^{\text{th}}$ antenna of receiver $k$, all the desired data streams at the $n^{\text{th}}$ antenna of receiver $k$ can be uniquely determined from $\tilde{Y}^{[k]}_{n}$. However, what we have received in the $n^{\text{th}}$ antenna of receiver $k$ is ${Y}^{[k]}_{n}$ which is a noisy version of $\tilde{Y}^{[k]}_{n}$. Therefore, to recover the desired data streams at the $n^{\text{th}}$ antenna of receiver $k$, we further require to accurately estimate $\tilde{Y}^{[k]}_{n}$ from ${Y}^{[k]}_{n}$. To this aim, let $\mu^{[k]}_n$ denote the rational dimension of the received pseudo-vectors at the $n^{\text{th}}$ antenna of receiver $k$. Apparently, $\mu^{[k]}_n\leq LMNK$. As we shall see shortly, if the rate control parameter $\nu$ in (\ref{Eq:LambdaSet}) is selected as:
\begin{equation}\label{Eq:nu}
\nu=\max_{k\in\xxK,n\in\xxN}\mu^{[k]}_n,
\end{equation}
then we would be able to identify $\tilde{Y}^{[k]}_{n}$ in ${Y}^{[k]}_{n}$ with high probability for all $k\in\xxK$ and all $n\in\xxN$.

Each user decodes its data on different receive antennas separately. In other words, there is no cooperation among receive antennas of each user. There are $ML$ desired data streams at the signal received by each antenna of every user. To decode each part, we treat the other parts as well as the interfering signals as i.i.d. noise and therefore as $\tau\rightarrow \infty$ the following rate is achievable for data stream $(m,l)$ of the signal received on the $n^{\text{th}}$ antenna of receiver $k$:
\begin{equation}    R^{[k]}_{mnl}=I(X^{[k]}_{mnl};Y^{[k]}_n)=H(X^{[k]}_{mnl})-H(X^{[k]}_{mnl}|Y^{[k]}_n),\,\,\,m\in\xxM,\,l\in\xxL,
\end{equation}
where for the notational simplicity, we omitted the time index $t$.
It is obvious that: \begin{align}H(X^{[k]}_{mnl})=\log_2|\Lambda^{[k]}_{mnl}(P,\epsilon)|\approx\frac{(1-\epsilon)}{2(\nu+\epsilon)}\log_2P+1.\end{align} In the following, we prove that if the modulation pseudo-vectors at all transmitters are selected such that the separability condition holds at all receive antennas of all receivers, then we \emph{almost always} have:
\begin{equation}\label{Eq:H(x|y)}
\limsup_{P\rightarrow\infty} H(X^{[k]}_{mnl}|Y^{[k]}_n)\leq c_0,\,\,\,\forall k\in\xxK,\,\forall m\in\xxM,\,\forall n\in\xxN,\,\forall l\in\xxL,
\end{equation}
 where $c_0$ is some constant independent of $P$. Consequently, user $k$ can almost always achieve $R^{[k]}_{mnl}=\frac{(1-\epsilon)}{2(\nu+\epsilon)}\log_2P+o(\log_2P)$ by decoding the $(m,l)$ data stream of its desired signal component on the $n^{\text{th}}$ receive antenna. Since there are $ML$ desired data streams in the signal received by the $n^{\text{th}}$ antenna  of user $k$ and since $\epsilon$ can be made arbitrarily small, it follows that $\dof\geq \frac{LMNK}{\nu}$.\\
Next, we show that (\ref{Eq:H(x|y)}) is valid under the above-mentioned conditions. Let
\begin{align}
\Theta^{[k]}_{n}(P,\epsilon)&\define\left\{\sum_{k'=1}^{K}\sum_{m=1}^M\sum_{n'=1}^N\sum_{l=1}^L h^{[kk']}_{nm}h^{[k'k']}_{n'm}\lambda^{[k']}_{mn'l}:\,\lambda^{[k']}_{mn'l}\in \Lambda^{[k']}_{mn'l}(P,\epsilon) \right\},\,\,k\in\xxK,\,n\in\xxN.
\end{align}
Note that $\Theta^{[k]}_{n}(P,\epsilon)$ is the support set of the random variable $\tilde{Y}^{[k]}_n$ which is the noise-free part of ${Y}^{[k]}_n$. We can estimate $\tilde{Y}^{[k]}_n$ from $Y^{[k]}_n$ using the following estimator:
\begin{equation}\label{Eq:estimator}
\widehat{\tilde{Y}^{[k]}_n}=\underset{\theta\in\Theta^{[k]}_{n}(P,\epsilon)}{\operatorname{argmin}}|Y^{[k]}_n-\theta|.
\end{equation}
  An error may occur using this estimation whenever the absolute value of the additive Gaussian noise $Z^{[k]}_n$ is greater than half of the minimum distance of the set $\Theta^{[k]}_{n}(P,\epsilon)$. That is
\begin{equation}\label{Eq:Pe}
    \text{Pr}\{\widehat{\tilde{Y}^{[k]}_n}\neq \tilde{Y}^{[k]}_n\}\leq \text{Pr}\left\{|Z^{[k]}_n|\geq\frac{d_{\text{min}}(\Theta^{[k]}_{n}(P,\epsilon))}{2}\right\}\leq 2\exp\left(-\frac{d^2_{\text{min}}(\Theta^{[k]}_{n}(P,\epsilon))}{8}\right),
\end{equation}
where the last inequality follows from the properties of Gaussian distribution. As we discussed earlier, if the separability condition holds at all antennas of all receivers, we can uniquely determine $X^{[k]}_{mnl}$ from $\tilde{Y}^{[k]}_n,\forall m\in\xxM\text{ and } \forall l\in\xxL$. Hence, $\text{Pr}\{\widehat{X}^{[k]}_{mnl}\neq X^{[k]}_{mnl}\}\leq \text{Pr}\{\widehat{{\tilde{Y}}^{[k]}_n}\neq \tilde{Y}^{[k]}_n\}.$ Therefore, we can upper-bound $H(X^{[k]}_{mnl}|Y^{[k]}_n)$ using the data processing and Fano's inequalities \cite{Etkin-Ordentlich}:
\begin{align}\label{Eq:fano}
    \nonumber H(X^{[k]}_{mnl}|Y^{[k]}_n)&\leq H(X^{[k]}_{mnl}|\hat{X}^{[k]}_{mnl})\leq 1+\text{Pr}\{\hat{X}^{[k]}_{mnl}\neq X^{[k]}_{mnl}\}\log_2(|\Lambda^{[k]}_{mnl}(P,\epsilon)|)\\
          &\leq 1+ 2\exp\left(-\frac{d^2_{\text{min}}(\Theta^{[k]}_{n}(P,\epsilon))}{8}\right)\times\left[\frac{(1-\epsilon)}{2(\nu+\epsilon)}\log_2P+1+o(1)\right]
\end{align}
Finally, we show that if $\nu$ is selected according to (\ref{Eq:nu}), then we almost always have $d_{\text{min}}(\Theta^{[k]}_{n}(P,\epsilon))\geq \varrho P^{\frac{\epsilon}{2}}$ for some constant $\varrho$. Accordingly, (\ref{Eq:H(x|y)}) follows from (\ref{Eq:fano}). If we select $\nu$ as in (\ref{Eq:nu}), then each $\theta^{[k]}_n\in\Theta^{[k]}_{n}(P,\epsilon)$ is a rational linear combination of at most $\nu$ rationally independent real numbers and therefore it can be expressed as:
\begin{align}
    \theta^{[k]}_n=\gamma P^{\frac{\nu-1+2\epsilon}{2(\nu+\epsilon)}}\sum_{i=1}^\nu \delta^{[k]}_{ni}T^{[k]}_{ni},
\end{align}
where $T^{[k]}_{ni}$'s, $i=1,\cdots,\nu$, represent $\nu$ rationally independent received pseudo-vectors\symbolfootnote[1]{Note that according to the separability condition, out of these $\nu$ rationally independent received pseudo-vectors, $ML$ ones are  $(h^{[kk]}_{nm})^2\,\omega^{[k]}_{mnl}$, $m\in\xxM$, $l\in\xxL$.} at the $n^{\text{th}}$ antenna of receiver $k$ and $\delta^{[k]}_{ni}$'s $,i=1,\cdots,\nu$ are the corresponding integer coefficients. Since at most $KM$ independent data streams may arrive along the same received pseudo-vector $T^{[k]}_{ni}$,  it follows that $|\delta^{[k]}_{ni}|\leq KMQ$. The minimum distance $d_{\text{min}}(\Theta^{[k]}_{n}(P,\epsilon))$ is the minimum value of $|\theta^{[k]}_n-{\theta'}^{[k]}_n|$, $\forall\theta^{[k]}_n\in\Theta^{[k]}_{n}(P,\epsilon)$,   $\forall{\theta'}^{[k]}_n\in\Theta^{[k]}_{n}(P,\epsilon)\setminus\theta^{[k]}_n$. The quantity $|\theta^{[k]}_n-{\theta'}^{[k]}_n|$ can be expressed as:
\begin{equation}\label{dmin}
    |\theta^{[k]}_n-{\theta'}^{[k]}_n|=\gamma P^{\frac{\nu-1+2\epsilon}{2(\nu+\epsilon)}}\left|\sum_{i=1}^\nu T^{[k]}_{ni}(\delta^{[k]}_{ni}-\delta'^{[k]}_{ni})\right|.
\end{equation}
 According to the Khintchine-Groshev Theorem, for every $\epsilon>0$ there exists some constant $c_1$ such that:
\begin{equation}
    \left|\sum_{i=1}^\nu T^{[k]}_{ni}(\delta^{[k]}_{ni}-\delta'^{[k]}_{ni})\right|\geq\frac{c_1}{(2KM Q)^{\nu-1+\epsilon}}
    \end{equation}
for almost all received pseudo-vectors $T^{[k]}_{ni}$'s, $i=1,\cdots,\nu$. Therefore, the minimum distance $d_{\text{min}}(\Theta^{[k]}_{n}(P,\epsilon))$ is lower-bounded by:
\begin{equation}
    d_{\text{min}}(\Theta^{[k]}_{n}(P,\epsilon))\geq\varrho P^{\frac{\nu-1+2\epsilon}{2(\nu+\epsilon)}}P^{-\frac{(1-\epsilon)(\nu-1+\epsilon)}
    {2(\nu+\epsilon)}}=\varrho P^{\frac{\epsilon}{2}}
    \end{equation}
for almost all received pseudo-vectors $T^{[k]}_{ni}$'s, $i=1,\cdots,\nu$, where $\varrho'=c_1\gamma (2KM)^{-(\nu-1+\epsilon)}$ is a constant independent of $P$.
Since the lower-bound on the minimum distance is obtained using the Khintchine-Groshev Theorem, we use the term ``almost always" in  statements concerning our achievability result.

So far, we established that for almost all modulation pseudo-vectors $h^{[kk]}_{mn}\omega^{[k]}_{mnl},\,k\in\xxK,\,m\in\xxM,\,n\in\xxN,\,l\in\xxL$ satisfying the separability condition at all antennas of all receivers, the proposed scheme can achieve $\frac{LMNK}{\nu}$ degrees of freedom where $\nu$ represents the maximum number of rationally independent received pseudo-vectors across all receive antennas of all users. In general, $\nu$ can be as large as $LMNK$ and therefore DoF strongly depends on the value of $\nu$. In the sequel, we show that if the modulation pseudo-vectors are properly selected according to the channel coefficients, the value of $\nu$ can approach $(M+N)L$, and consequently, $K\frac{MN}{M+N}$ degrees of freedom is almost always achievable. As mentioned earlier, reducing $\nu$ by an appropriate selection of modulation pseudo-vectors is counterpart to the alignment condition in signal space alignment.
We define $\mathcal{H}^{[k]}_m$ as the set of channel coefficients from the $m^{\text{th}}$ antenna of user $k$ to all receive antennas of different users. That is:
$$\mathcal{H}^{[k]}_m\define\{h^{[1k]}_{1m},h^{[1k]}_{2m},\cdots,h^{[1k]}_{Nm},h^{[2k]}_{1m},h^{[2k]}_{2m},\cdots,h^{[2k]}_{Nm},\cdots,
h^{[Kk]}_{1m},h^{[Kk]}_{2m},\cdots,h^{[Kk]}_{Nm}\}.$$
Note that $|\mathcal{H}^{[k]}_m|=KN,\,\forall k\in\xxK,\,\forall m\in\xxM$. For each $n\in\xxN$, we define $E_{n}$ as:
\begin{align}\label{Eq:En}
E_{n}\define\bigcup_{k=1}^{K}\bigcup_{m=1}^{M}\left[h^{[kk]}_{nm}.(\mathcal{H}^{[k]}_m\setminus h^{[kk]}_{nm})\right].
\end{align}
Note that each element of $E_n$ is the product of two channel coefficients. That is if $e\in E_n$, then $e$ can be represented as $h^{[kk]}_{nm}h^{[k'k]}_{n'm}$ for some $k\in\xxK$, $k'\in\xxK$, $m\in\xxM$, $n'\in\xxN$ where $(k,n)\neq (k',n')$.
%
One can verify that $|E_{n}|=KM(KN-1),\,\forall n\in\xxN$. For a positive integer $\Gamma$ and for each $m\in\xxM$, $n\in\xxN$, $k\in\xxK$, we select $\Omega^{[k]}_{mn}$ as:
\begin{align}\label{Eq:Omega}
\Omega^{[k]}_{mn}=\left\{\prod_{i=1}^{|E_{n}|}e_i^{s_i}\,:\,e_i\in E_{n},\,s_i\in\{0,1,\cdots,\psi^{[k]}_{mn}(e_i)\}\right\},
\end{align}
where $\psi^{[k]}_{mn}(\cdot)$ are functions described by:
\begin{align}\label{Eq:realDirectionExponents}
    \psi^{[k]}_{mn}(e)=\left\{
    \begin{array}{ll}
      \Gamma-1, & \text{if } e\in h^{[kk]}_{nm}.(\mathcal{H}^{[k]}_m\setminus h^{[kk]}_{nm})\\
      \Gamma, & \text{Otherwise}
    \end{array}\right.
    .
\end{align}
We claim that if the real numbers $\omega^{[k]}_{mnl}$ are selected from $\Omega^{[k]}_{mn}$ in (\ref{Eq:Omega}), then the separability condition holds at all antennas of all receivers and moreover $\nu$ can approach $(M+N)L$.
First, we notice that elements of $\Omega^{[k]}_{mn}$ are different monomials in the variables $e_i$'s and therefore they are almost always linearly independent. From (\ref{Eq:En}), (\ref{Eq:Omega}), and (\ref{Eq:realDirectionExponents}), one can verify that the number of modulation pseudo-vectors, $L$, which is equal to the cardinality of $\Omega^{[k]}_{mn}$, is given by
\begin{align}\label{Eq:L}L=\Gamma^{KN-1}(\Gamma+1)^{(KM-1)(KN-1)}.\end{align}
Next, consider the received signal at the $n^{\text{th}}$ antenna of receiver $k$ at time index $t$. From (\ref{Eq:Ykn2}), we see that:
\begin{itemize}
	\item Received pseudo-vectors corresponding to the desired component of $Y^{[k]}_{n}(t)$ are the elements of $\bigcup_{m=1}^{M}(h_{nm}^{[kk]})^2\cdot \Omega^{[k]}_{mn}$.
	\item Received pseudo-vectors corresponding to the self-interference component of $Y^{[k]}_{n}(t)$ are the elements of
	$\mathcal{B}^{[k]}_n\define\bigcup_{m=1}^{M}\bigcup_{\substack{n'=1\\n'\neq n}}^{N}\left[h_{nm}^{[kk]}h_{n'm}^{[kk]}\cdot \Omega^{[k]}_{mn'}\right]$.
	\item Received pseudo-vectors corresponding to the multi-user interference component of $Y^{[k]}_{n}(t)$ are the elements of
	$\mathcal{G}^{[k]}_{n}\define\bigcup_{\substack{k'=1\\k'\neq k}}^{K}\bigcup_{m=1}^{M}\bigcup_{n'=1}^{N}\left[h_{nm}^{[kk']}h_{n'm}^{[k'k']}\cdot \Omega^{[k']}_{mn'}\right]$.
\end{itemize}

 Since $(h^{[kk]}_{nm})^2\notin E_n$, $\forall k\in\xxK,\,\forall m\in\xxM,\,\forall n\in\xxN$, it follows that the received pseudo-vectors corresponding to the desired component can not be expressed as rational linear combinations of the other received pseudo-vectors and therefore the separability condition holds at all antennas of all receivers. We then notice that:
 \begin{align}\label{Eq:monomial}
 \begin{split}
  h_{nm}^{[kk]}h_{n'm}^{[kk]}&\in E_{n'}, \quad\forall m\in\xxM,\,n'\neq n\\
  h_{nm}^{[kk']}h_{n'm}^{[k'k']}&\in E_{n'}, \quad\forall m\in\xxM,\,k'\neq k
  \end{split} .
  \end{align}
  Since each element of $\Omega^{[k]}_{mn'}$, $n'\neq n$, is a monomial in the variables $e'_i$'s where $e'_i\in E_{n'}$, and because of (\ref{Eq:monomial}), each element of $\bigcup_{m=1}^{M}\left[h_{nm}^{[kk]}h_{n'm}^{[kk]}\cdot \Omega^{[k]}_{mn'}\right]$ is again a monomial in $e'_i$'s with a degree at most $\Gamma$ for each variable. Similarly, since each element of $\Omega^{[k']}_{mn'}$, $k'\neq k$ is a monomial in $e'_i$'s where $e'_i\in E_{n'}$, and because of (\ref{Eq:monomial}), each element of $\bigcup_{\substack{k'=1\\k'\neq k}}^{K}\bigcup_{m=1}^{M}\left[h_{nm}^{[kk']}h_{n'm}^{[k'k']}\cdot \Omega^{[k']}_{mn'}\right]$ is again a monomial in $e'_i$'s with a degree at most $\Gamma$ for each variable. Hence,
  \begin{align}
  \dim(\mathcal{B}^{[k]}_n\bigcup\mathcal{G}^{[k]}_{n})\leq N(\Gamma+1)^{KM(KN-1)}.
  \end{align}
  Therefore,
  \begin{align}\mu^{[k]}_{n}\leq  ML+N(\Gamma+1)^{KM(KN-1)}.\end{align}
  Recall that $\mu^{[k]}_{n}$ is the rational dimension of the received pseudo-vectors at the $n^{\text{th}}$ antenna of receiver $k$. We then have:
\begin{align}\label{Eq:nuFinal}\nu\leq ML+N(\Gamma+1)^{KM(KN-1)}.\end{align}
Therefore, from (\ref{Eq:L}) and (\ref{Eq:nuFinal}) the achievable DoF is given by:
$$\dofl= \frac{KMN\Gamma^{KN-1}(\Gamma+1)^{(KM-1)(KN-1)}}{M\Gamma^{KN-1}(\Gamma+1)^{(KM-1)(KN-1)}+N(\Gamma+1)^{KM(KN-1)}}.$$
Noting that $\Gamma$ is an arbitrary integer, as $\Gamma\rightarrow \infty$, the achievable DoF tends to $K\frac{MN}{M+N}$.
\section{Conclusions}\label{sec conclusions}
In this paper, we obtained new results for the DoF
of the fully connected constant MIMO interference channel. We showed how
real interference alignment can be used to achieve
a higher DoF for MIMO interference channel. We also introduced
a new upper-bound on the DoF for a MIMO interference
channel, which coincides with our achievable DoF
when the number of users is larger than some threshold, which
depends on the number of transmit and receive antennas.
\appendices
\section{Calculating the normalizing constant $\gamma$ in (\ref{Eq:LambdaSet})}\label{App:PowerConstraint}
The average transmit power of user $k$ can be calculated as follows:
\begin{align}\label{Eq:AppPower1}
\sum_{m=1}^M\mathbb{E}\left[(X^{[k]}_{m})^2\right]=\sum_{m=1}^M\sum_{n=1}^{N}(h^{[kk]}_{nm})^2\mathbb{E}\left[(X^{[k]}_{mn})^2\right]=\sum_{m=1}^M\sum_{n=1}^{N}\sum_{l=1}^{L}(h^{[kk]}_{nm})^2 \mathbb{E}\left[(X^{[k]}_{mnl})^2\right].
\end{align}
On the other hand, since $X^{[k]}_{mnl}$ is uniformly distributed over $\Lambda^{[k]}_{mnl}(P,\epsilon)$, it follows that
\begin{align}
\mathbb{E}\left[(X^{[k]}_{mnl})^2\right]=\frac{1}{|\Lambda^{[k]}_{mnl}(P,\epsilon)|}\sum_{x\in\Lambda^{[k]}_{mnl}(P,\epsilon)}x^2,
\end{align}
where $|\Lambda^{[k]}_{mnl}(P,\epsilon)|$ denotes the size of the set $\Lambda^{[k]}_{mnl}(P,\epsilon)$ which is equal to $2Q+1$. Therefore,
\begin{align}\label{Eq:AppPower2}
\mathbb{E}\left[(X^{[k]}_{mnl})^2\right]=\frac{\gamma^2 P^{\frac{
\nu-1+2\epsilon}{\nu+\epsilon}}\,\left(\omega^{[k]}_{mnl}\right)^2}{2Q+1}\sum_{q=-Q}^{Q}q^2=\gamma^2 P^{\frac{
\nu-1+2\epsilon}{\nu+\epsilon}}\,\left(\omega^{[k]}_{mnl}\right)^2 \frac{Q(Q+1)}{3}.
\end{align}
Substituting (\ref{Eq:AppPower2}) in (\ref{Eq:AppPower1}) and noting that $Q(Q+1)\approx P^{\frac{
1-\epsilon}{\nu+\epsilon}}$ for large values of $P$, we obtain:
\begin{align}
\sum_{m=1}^M\mathbb{E}\left[(X^{[k]}_{m})^2\right]\approx\frac{1}{3}\gamma^2 P\sum_{m=1}^M\sum_{n=1}^{N}\sum_{l=1}^{L}(h^{[kk]}_{nm}\omega^{[k]}_{mnl})^2.
\end{align}
Therefore, the power constraint $P$ at all transmitters is satisfied if
$$\gamma^2=\underset{k\in\xxK}{\operatorname{min}}\frac{3}{\sum_{m=1}^M\sum_{n=1}^{N}\sum_{l=1}^{L}(h^{[kk]}_{nm}\omega^{[k]}_{mnl})^2}.$$
\section{Proof of (\ref{Eq:2user upperbound2})}\label{App:Upper-boundProof}
In this appendix, we prove that
\begin{equation}\label{Eq:2user upperbound3}
   J(W_1M,W_2M,W_1N,W_2N)\leq \max\{\max(M,N)W_{\text{min}},
   \min(M,N)W_{\text{max}}\},
\end{equation}
where $W_{\text{min}}=\min(W_1,W_2)$ and $W_{\text{max}}=\max(W_1,W_2)$. First, note that
\begin{align}\label{App: upper1}
\begin{array}{rl}
J(W_1M,W_2M,W_1N,W_2N)&=\min\{WM,WN,\max(W_1M,W_2N),\max(W_2M,W_1N)\}\\
&\leq \min\{\max(W_1M,W_2N),\max(W_2M,W_1N)\}.
\end{array}
\end{align}
Due to the symmetry, without loss of generality, we prove (\ref{Eq:2user upperbound3}) for the case of $M\geq N$. We consider two cases:
\begin{enumerate}
  \item $W_1\geq W_2$ \\
  In this case, $\max(W_1M,W_2N)=W_1M$. To evaluate (\ref{App: upper1}), we  differentiate between two cases:
  \begin{itemize}
    \item $W_1N\geq W_2M$\\
    In this case, $\max(W_2M,W_1N)=W_1N$. Therefore, (\ref{App: upper1})  reduces to:
  \begin{align*}
J(W_1M,W_2M,W_1N,W_2N)&\leq \min\{W_1M,W_1N\}=W_1N\\
&=\max\{W_1N,W_2M\}=\max\{W_{\text{max}}N,W_{\text{min}}M\}.
\end{align*}
    \item $W_1N< W_2M$\\
    In this case $\max(W_2M,W_1N)=W_2M$. Therefore, (\ref{App: upper1})  reduces to:
  \begin{align*}
J(W_1M,W_2M,W_1N,W_2N)&\leq \min\{W_1M,W_2M\}=W_2M\\&=\max\{W_1N,W_2M\}=\max\{W_{\text{max}}N,W_{\text{min}}M\}.
\end{align*}
  \end{itemize}
  \item $W_1< W_2$ \\
  In this case, $\max(W_2M,W_1N)=W_2M$. To evaluate (\ref{App: upper1}), we  again differentiate between two cases:
  \begin{itemize}
    \item $W_1M\geq W_2N$\\
    In this case, $\max(W_1M,W_2N)=W_1M$. Therefore, (\ref{App: upper1})  reduces to:
  \begin{align*}
J(W_1M,W_2M,W_1N,W_2N)&\leq \min\{W_1M,W_2M\}=W_1M\\
&=\max\{W_1M,W_2N\}=\max\{W_{\text{max}}N,W_{\text{min}}M\}.
\end{align*}
    \item $W_1M< W_2N$\\
    In this case, $\max(W_1M,W_2N)=W_2N$. Therefore, (\ref{App: upper1})  reduces to:
  \begin{align*}
J(W_1M,W_2M,W_1N,W_2N)&\leq \min\{W_2N,W_2M\}=W_2N\\
&=\max\{W_1M,W_2N\}=\max\{W_{\text{max}}N,W_{\text{min}}M\}.
\end{align*}
  \end{itemize}
 This completes the proof.
\end{enumerate}
\section{The closest rational neighbors of a real number with denominator at most $K$}\label{App:RationalApprox}
In this appendix, we study how closely a real number can be approximated
by rational numbers that have a given bound on the size of their denominators. Specifically, for a real number $\alpha$ and a positive integer $K$, we are looking for two rational numbers $\alpha^-$ and $\alpha^+$ such that $\alpha^-\leq \alpha\leq \alpha^+$ and moreover $\alpha^-$ and $\alpha^+$ are closer to $\alpha$ than any other rational number with denominator at most $K$. Given $\alpha$ and $K$, there is an elegant method to find the rationals $\alpha^-$ and $\alpha^+$ using the so called Farey sequence\cite{Mathematics}. A Farey sequence of order $N$ consists of all irreducible fractions from $[0, 1]$ with denominator not
exceeding $N$, arranged in order of increasing
magnitude. The Farey sequence of order $N$ will be denoted by $\sF_N$.
For example $\sF_5 = \{\frac{0}{1}, \frac{1}{5}, \frac{1}{4}, \frac{1}{3}, \frac{2}{5}, \frac{1}{2},\frac{3}{5}, \frac{2}{3}, \frac{3}{4}, \frac{4}{5}, \frac{1}{1}\}$. 

Suppose that $\alpha\in[0, 1)$ is a given real number, and the goal is to calculate the closest rational neighbors of $\alpha$ with denominator not exceeding a given positive integer $K$. To do this, we need to find the place of $\alpha$ in the sequence $\sF_K$. If $\alpha\in \sF_k$, then $\alpha^-=\alpha^+=\alpha$. If $\alpha\notin \sF_k$, then we can find its closest rationals $\alpha^-$ and $\alpha^+$ by:
\begin{align}\label{Eq:FareyOpt}
\alpha^-=\max_{\substack{q\in\sF_K\\q<\alpha}}q,\qquad
\alpha^+=\min_{\substack{q\in\sF_K\\q>\alpha}}q.
\end{align}
 For example, the closest rational neighbors of $\alpha=\sqrt{2}-1$ with denominator not exceeding 5 are $\alpha^-=\frac{2}{5}$ and $\alpha^+=\frac{1}{2}$. In this method, for a given $K$, we first need to construct the sequence $\sF_K$ and then solve the optimization problem in (\ref{Eq:FareyOpt}). 
 Lemma 1 provides an alternative approach to find the closest rational neighbors of a given real number $\alpha$ with denominator at most $K$ without the help of Farey sequence.

\begin{proof}[Proof of Lemma 1]
To prove (\ref{eq:neigbours1}), let us assume that $\max_{\substack{n\in\{1,\cdots,K\}}}\frac{\lfloor n\alpha\rfloor}{n}=\frac{\lfloor n_0\alpha\rfloor}{n_0}$ for some $n_0\in\{1,\cdots,K\}$. Note that $\frac{\lfloor n_0\alpha\rfloor}{n_0}\leq \alpha$ and $\frac{\lfloor n_0\alpha\rfloor}{n_0}\in\sF_K$. We claim that among all fractions in $\sF_K$ that are less than $\alpha$, the fraction $\frac{\lfloor n_0\alpha\rfloor}{n_0}$ is the closest to $\alpha$. We prove our claim by contradiction. Assume we can find a fraction $\frac{p}{q},\,(p,q)=1$ such that $\frac{p}{q}\in \sF_K$ and $\frac{\lfloor n_0\alpha\rfloor}{n_0}<\frac{p}{q}\leq \alpha$. It then follows that:
\begin{equation}\label{Eq:ProofThmContra1}
p\leq q\alpha.
 \end{equation}
 On the other hand, since $q\leq K$, it follows that $\frac{\lfloor  q\alpha\rfloor}{q}\leq \frac{\lfloor n_0\alpha\rfloor}{n_0}$ and since $\frac{\lfloor n_0\alpha\rfloor}{n_0}<\frac{p}{q}$ it follows that
 \begin{equation}\label{Eq:ProofThmContra2}
p>\lfloor q\alpha\rfloor .
 \end{equation}
 Combining (\ref{Eq:ProofThmContra1}) and (\ref{Eq:ProofThmContra2}), we have $\lfloor q\alpha \rfloor<p\leq q\alpha$ which is a contradiction because $p$ is an integer. We can prove (\ref{eq:neigbours1}) by a similar argument.
\end{proof}


\end{document}